\documentclass[conference]{IEEEtran}

\IEEEoverridecommandlockouts
\usepackage{graphicx}
\newcommand{\code}[1]{\texttt{#1}}

\newcommand{\url}[1]{#1}
\newcommand{\hide}[1]{}
\newcommand{\anon}[2]{#1}

\newcommand{\pic}[3]{\raisebox{#3}{\includegraphics[width=#2]{#1}}}

\begin{document}

\title{Using Non-Verbal Expressions\\ as a Tool in Naming Research%
\thanks{\anon{This research was supported by the ISRAEL SCIENCE FOUNDATION (grant no.\ 832/18).}{}}
}

\author{\anon{\IEEEauthorblockN{Omer Regev$^*$\thanks{$^*$ Authors contributed equally.} ~~~~~ Michael Soloveitchik$^*$ ~~~~~ Dror G. Feitelson}
\IEEEauthorblockA{Department of Computer Science\\
The Hebrew University of Jerusalem, 91904 Jerusalem, Israel}}
{Authors Anonymized}}

\maketitle

\begin{abstract}
Variable and function names are extremely important for program comprehension.
It is therefore also important to study how developers select names.
But controlled experiments on naming are hindered by the need to describe to experimental subjects what it is they need to name.
Words appearing in these descriptions may then find their way into the names, leading to a bias in the results.
We suggest that this problem can be alleviated by using emojis or other small graphics in lieu of key words in the descriptions.
A replication of previous work on naming, this time including such emojis and graphics, indeed led to a more diverse and less biased choice of words in the names than when using English descriptions.
\end{abstract}
\begin{IEEEkeywords}
Variable and function naming, experimental methodology, accessibility bias
\end{IEEEkeywords}

\vspace{1mm}\begin{raggedleft}\it
There are only two hard things in Computer Science:\\
cache invalidation and naming things.\\
-- Phil Karlton\\[5mm]
\end{raggedleft}

\section{Introduction}
%=====================

It is well-known that names of variables and functions play a major role in programs’ source code.
In large open source projects about a third of the tokens are identifiers, and they account for about two thirds of the characters in the source code \cite{deissenboeck05}.
However, the importance of names is not based only on their volume.
Their main importance is that they serve as implicit documentation, and convey the code’s intended functionality \cite{salviulo14}.
In fact, sometimes names are the only documentation.
This is even advocated as part of the ``clean code'' approach, which states that ``if a name requires a comment, then the name does not reveal its intent'' \cite{martin:clean}.

The importance of variable and function names for comprehension has stimulated a considerable amount of research on this topic.
One favorite topic of research has been how the length of variable names affects comprehension, and especially the difference between using full words and abbreviations \cite{newman19,hofmeister19,lawrie06,scanniello13,schankin18,takang96}.
Research on actual usage has shown that names are getting longer \cite{holzmann16}, and that longer names are especially characteristic of ``engineered'' code \cite{lemos20}, but that single-letter names are also being used \cite{beniamini17}.
The strain that longer names place on memory has also been studied \cite{binkley09c}.
Another topic has been naming style \cite{binkley13}, for example using camelCase or under\_score \cite{binkley09b} and the effect of naming conventions \cite{butler15}.

Attention has also been given to the quality of names, and to how naming relates to the cognitive processes involved in program comprehension \cite{raychev15,liblit06}.
Several studies have dealt with bad names and their ill-effects on comprehension \cite{arnaoudova16,avidan17,liu19}, including due to cognitive load \cite{fakhoury19b}.
Bad names have also been linked with low code quality in general \cite{butler10,aman15}.
When developers encounter bad names they may change them (a form of refactoring), so it is interesting to see what replacements they choose \cite{arnaoudova14}.
There has also been work on suggesting names automatically, e.g.\ based on machine learning of code \cite{allamanis14b,raychev15}.

Finally, several studies have suggested how practitioners may create better names.
This is usually based on either (or both) of two methods.
One is controlling the vocabulary used, so as to avoid synonyms and homonyms and reduce ambiguity  \cite{deissenboeck05,binkley15,caprile00,fakhoury18,lawrie07b}.
The other is to formulate rules for structuring names in a consistent manner \cite{binkley11,caprile00}.

The majority of all these studies have used one of two empirical methods:
either controlled experiments where subjects need to deal with code containing different variable names, or repository mining to see what names were used in real code.
But there have been very few studies based on experiments in which developers are presented with a situation and are actually required to select names.
This is unfortunate, as naming is acknowledged to be hard, so it deserves to be studied directly.

A major problem in studying spontaneous naming is that the context needs to be explained to the experimental subjects.
But the description of the context, and the question regarding the naming of variables in this context, necessarily use words.
Being exposed to these words makes them more accessible, so subjects will tend to use words from the description and question in the names they create.
Therefore, providing the description undermines the spontaneity we are trying to characterize.

This accessibility bias was demonstrated in a recent study about naming by Feitelson et al.\ \cite{feitelson:names}.
In this study several programming scenarios were drawn up, and programmers were required to name variables and functions that were expected to be used in them.
As expected, the selected names were strongly influenced by the words used in the scenario descriptions and the naming questions.
The suggested solution was to use bilingual experimental subjects, and present the naming context and questions in a language other then English --- in this case, in Hebrew.
As English is typically the language used in programming, this creates a separation between the experimental materials and the names.
The results were that indeed the bias was much reduced when the descriptions and the questions were given in Hebrew, and experimental subjects used a much wider variety of words in the names they suggested.

However, This approach did not solve the problem completely, and suffered from several drawbacks.
First, even the Hebrew descriptions caused some bias on the chosen names, which was manifested in the appearance of transcribed Hebrew words or their direct translations into English.
Second, this research method can be applied only with developers that are bilingual, something that is not true for many developers and in particular is not true to most developers who live in English speaking countries.
It therefore has limited applicability.
Third, this approach is restricted only to subjects who speak the second language (e.g.\ Hebrew) fluently.
This implies a reduced representativeness: the subjects represent second-language-speaking (e.g.\ Hebrew-speaking) developers only.

As an alternative we suggest using a graphical notation rather than a foreign language.
Specifically, we suggest that key words in the scenario description and questions be replace by emojis or other small graphics.
This could help reduce the accessibility bias by providing a more direct indication of the concept, without using any specific word as an intermediary.
And emojis are universally known and understood, making the methodology applicable anywhere including with developers who only speak English.

To check this idea we performed a replication of the Feitelson et al.\ study.
We used the same scenarios and questions as in the original study, but expressed them with the aid of emojis and small graphics.

\section{Research Questions}
%===========================

The general context of our work is the experimental methodology used is naming studies, where subjects are required to choose names for variables in certain programming scenarios.
In such studies the description of the scenario might bias the name choice.
Our hypothesis is that by using emojis in the description, instead of explicit words, the bias can be much reduced.
The goal of our work is therefore to assess whether non-verbal expressions such as emojis can serve as a tool that improves research on choosing names.
This goal will be achieved by answering the following concrete research questions:
\renewcommand{\theenumi}{RQ\arabic{enumi}}
\begin{enumerate}[leftmargin=10mm]
\item\label{rq:emojis} Are emojis expressive enough to describe programming scenarios at a level comparable to that of verbal descriptions?
Are the concepts represented by the emojis understood?
\item\label{rq:accessibility} Does using emojis to describe programming scenarios reduce or eliminate the accessibility bias in naming which exists when using verbal descriptions?
\end{enumerate}
\renewcommand{\theenumi}{\arabic{enumi}}

\section{Methodology}
%====================

We propose to alleviate the problem of accessibility bias caused by using verbal descriptions by using emojis.
To evaluate this solution, we replicate the study of Feitelson et al.:
we translate the descriptions that were given in that study into emoji-based descriptions, use (almost) exactly the same experimental procedure, and analyze the results in comparison to the previous results.

\subsection{Sources for Emojis}

It is not widely known that emojis constitute a font, and are part of the Unicode character set.
Defining new ones is controlled by the Unicode Consortium.
Interestingly, one of the criteria for new emoji is ``does the emoji add to what can be said using emoji''.
At the time of writing, 3521 emojis are defined.
Many of these are encoded as a sequence of more than one Unicode characters, some of which are modifiers of the basic form (e.g.\ to change skin color).
The full list of emojis version 13.1, specifically not including skin tone variations, contains 1816 symbols.

Given that emojis are a font, the actual graphical image used for each one may vary by platform.
For example, the image for ``nerd face'' (code 1F9D0) can be any of
\pic{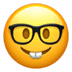}{6mm}{-1mm}
\pic{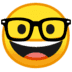}{6mm}{-1mm}
\pic{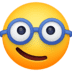}{6mm}{-1mm}
\pic{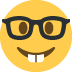}{6mm}{-1mm}
\pic{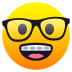}{6mm}{-1mm}
\pic{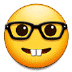}{6mm}{-1mm}
and more.
In particular, this means that the recipient of a message may see a variation on what the sender intended.
We therefore used actual embedded images rather than coded emojis in the experiments.

An obvious source for images of emojis is the full table of Unicode emojis (\url{https://unicode.org/emoji/charts/full-emoji-list.html}).
Another good site is \url{https://emojipedia.org/}.
And of course one can simply look for various icons on Google if there is no suitable emoji.
Some of the images we used are indeed not formally emojis, but we call all of them ``emoji'' for short.

\subsection{Translating Scenario Descriptions to Emojis}
\label{sect:trans}

The translation of the descriptions using emojis was done according to the following procedure.
The description here is slightly refined based on what we learned from the experiment.

\begin{enumerate}
\item Given a scenario and questions regarding the naming of variables and functions that are expected to be used, make a list of these target identifiers.
The target identifiers are the focus of the study, and we want to ensure that the description and questions do not suggest specific words for naming them.

\item Identify the key words in the description and questions (which can be nouns, verbs, adjectives, etc.) that describe the target identifiers.
These are words for which one of the followings holds:
\begin{enumerate}
    \item They appear as a direct description of a target identifier in the description of the scenario or in a question about a variable.
    Example: the words ``X coordinate'' in the question ``name the variable holding the X coordinate of a location on the grid''.
    
    \item They might reflect on a target identifier in some way:
    \begin{enumerate}
        \item A synonym of a word that describes a target variable.
        Example: ``horizontal displacement'' instead of ``X coordinate''.
        
        \item The word is often used in conjunction with a target word, and might identify it.
        Exampe: if we want to ask about the time that the bus will arrive at the bus stop, using ``\pic{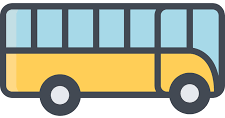}{7mm}{-1mm} stop'' is too explicit, and will direct subjects to interpret \pic{bus.png}{7mm}{-1mm} specifically as ``bus'' as opposed to any other possible form of transport.
        So we need to avoid using ``stop'' here.
        
        \item The word is often used in the same context as the target word, and might cause a bias.
        Example: in the context of mines in the well-known minesweeper game, we need to avoid not only ``mine'' but also ``bomb'', ``explode'', etc.
        In particular, if suspicious locations on the game board are marked by a \pic{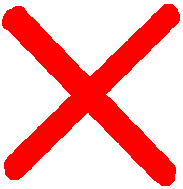}{4mm}{-1mm}, any of these words may bias subjects towards a specific interpretation as the location of a mine as opposed to just marking a location or perhaps some level of danger.
    \end{enumerate}
\end{enumerate}
These are all words we will want to replace by emojis.

\item Expand each word with its synonyms.
For each word in the list write its general meaning and related words.
For instance, if the word in the description is ``salary'', we can add ``wages'', ``payment'', ``money'', etc.
This step helps us in two ways:
\begin{enumerate}
    \item Generalizing the notion of the word, which will help us look for an emoji that reflects the general meaning, so the experimental subjects will have to think of a specific word by themselves.
    For example, when we expand ``salary'' with ``wages'', ``payment'', ``money'', etc., we deduce that the general idea we need to represent is ``money transfer''.
    Then an emoji which conveys this notion, e.g.\ \pic{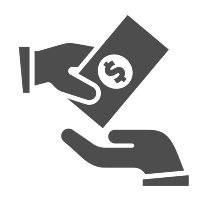}{8mm}{-3mm}, is general enough to allow the subjects to think of ``salary'', ``wages'', ``payment'', etc.
    
    \item Helping us beware of emojis that will remove bias of the original word, but cause a bias to a different specific word.
    For instance, in our experiment, we wished to replace the word ``benefits'' in a credit-card scenario with a more general indication, and used \pic{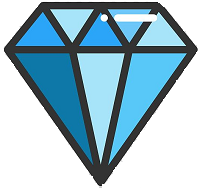}{6mm}{-2mm}.
    But then there was a new bias for the word ``diamonds''.
\end{enumerate}

\item Map words to emojis: for each word in the expanded list (original key words + synonyms), find 2--3 emojis which  reflect its general meaning or a close one (as explained above).
We look for generality --- emojis that explain the essence but not too much.
At the same time we need to be aware of words that will cause bias (e.g.\ the ``benefits''-``diamonds'' case).

\item Rewrite each sentence by replacing key words with emojis from the
emojis list.
An interesting issue is whether to stick to a fixed translation for each key word.
We preferred to alternate between different emojis that represent the same word.
Iterating can ensure the semantics are diverse enough, which makes the text less verbal and more abstract.
In places where a suggested emoji might have a too specific meaning, one can use a combination of 2 or 3 emojis as a whole word.

\item Make adaptations to the original descriptions wherever no reasonable emoji exists, or the meaning will not be clear with emojis.
Example: in one of the scenarios there was a variable representing the hourly wage of workers in a candy factory.
We represented this by \pic{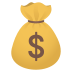}{5mm}{-1mm}-\pic{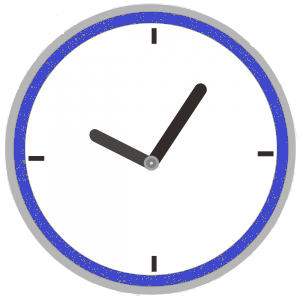}{5mm}{-1mm}.
However, using \pic{clock.png}{5mm}{-1mm} may cause confusion, as it appears more often as representing ``time'' or ``clock'' rather than ``hours''.
A possible alternative to consider is to replace hourly wage with daily wage, as represented by \pic{money.png}{5mm}{-1mm}-\pic{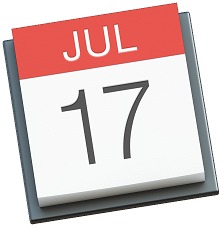}{6mm}{-2mm}.
\end{enumerate}

\begin{figure*}
    \centering
\fbox{~\parbox{0.9\textwidth}{
In a large \pic{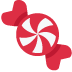}{5mm}{-1mm}\pic{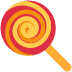}{5mm}{-1mm} company, \pic{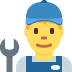}{6mm}{-1mm} earn \pic{money.png}{5mm}{-1mm} per \pic{clock.png}{5mm}{-1mm}.
Every \pic{worker.png}{6mm}{-1mm} has a fixed \pic{money.png}{5mm}{-1mm}-\pic{clock.png}{5mm}{-1mm} value.

Purim is right around the corner and causes an increased demand for \pic{candy1.png}{5mm}{-1mm}\pic{candy2.png}{5mm}{-1mm}.
To overcome this the company encourages \pic{worker.png}{6mm}{-1mm}\pic{worker.png}{6mm}{-1mm} to do more:
\begin{itemize}
    \item A \pic{worker.png}{6mm}{-1mm} usually does 45 \pic{clock.png}{5mm}{-1mm} per \pic{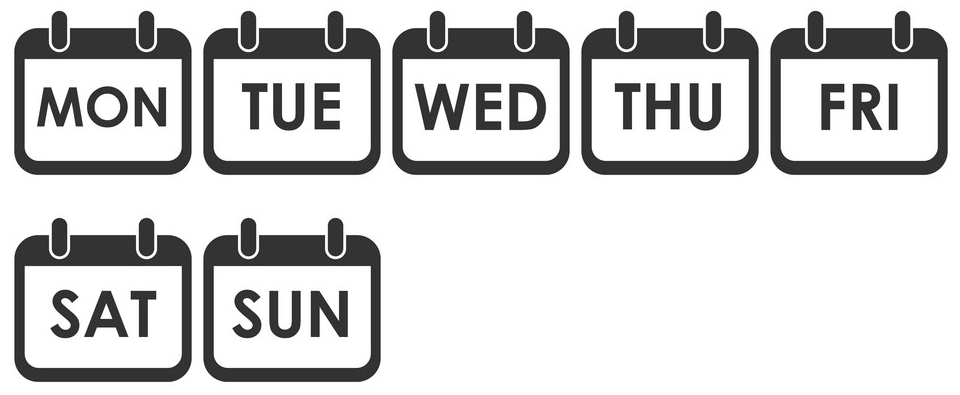}{15mm}{-1mm}
    \item After 45 \pic{clock.png}{5mm}{-1mm}, the \pic{money.png}{5mm}{-1mm}-\pic{clock.png}{5mm}{-1mm} for the \pic{worker.png}{6mm}{-1mm} increases by \$10.
\end{itemize}
To implement this some variables were added:
\begin{enumerate}
    \item A constant containing the value 45
    \item A variable for the \pic{money.png}{5mm}{-1mm}-\pic{clock.png}{5mm}{-1mm} after 45 \pic{clock.png}{5mm}{-1mm} pass
\end{enumerate}
Name these variables.
\begin{enumerate}
    \item Name variable: A constant containing the value 45.
    \item Name variable: A variable for the \pic{money.png}{5mm}{-1mm}-\pic{clock.png}{5mm}{-1mm} after (1) passes.
\end{enumerate}
}~}\\[1mm]
\fbox{~\parbox{0.9\textwidth}{
In a large chewing gum company, workers earn hourly (NIS). Every employee has a fixed hourly wage value.

Purim is right around the corner and the Mishlochei Manot cause an increased demand for chewing gum. To overcome this the factory manager encourages employees to work overtime as follows:
\begin{itemize}
    \item A full-time position requires 45 weekly work hours.
    \item After 45 weekly work hours, the hourly wage for the employee increases by 10 ILS.
\end{itemize}
To implement this some variables were added:
\begin{itemize}
    \item A constant containing the value 45
    \item A variable for the hourly wage during overtime pay
\end{itemize}
\begin{enumerate}
    \item Name the constant containing the value 45.
    \item Name the variable for the hourly wage during overtime pay.
\end{enumerate}
}~}
    \caption{\sl Example scenario and questions from the experiment, and the original version from which it was derived.
    Note changes in wording where there was no suitable emoji, e.g.\ ``to do more'' instead of ``work overtime''.}
    \label{fig:Q-ex}
\end{figure*}

An example of a scenario description followed by two questions from the experiment, after translation to emojis, is shown in Fig.\ \ref{fig:Q-ex}.
Note that several changes were made to better fit the emoji vocabulary.
First, bubble-gum was replaced by \pic{candy1.png}{5mm}{-1mm}\pic{candy2.png}{5mm}{-1mm}.
Second, the reference to the company manager was removed as we did not find a suitable emoji for ``company manager''.
Likewise, referring to a 45-hour week explicitly as a ``full-time position'' was also removed.
Finally, the explicit term ``work overtime'' was replaced by the more general ``to do more''.

The fact that we managed to present all the questions using emojis provides part of the answer to \ref{rq:emojis}: apparently emojis are expressive enough, provided some adjustments are made.
But we still need to see if they are understood correctly.
This is done in Section \ref{sect:understand}.

\subsection{Execution of the Experiment}

The execution of the experiment comprised of the following steps:
\begin{enumerate}
    \item Taking the same descriptions used in the research conducted by Feitelson et al.\ \cite{feitelson:names}, and translating them into emoji syntax according to the steps described above in Section \ref{sect:trans}.
    We made sure that all key words were translated into emojis.

    \item Creating a questionnaire using the Google Forms platform.
    While the original study included 11 scenarios, we only used six.
    Three were excluded because they did not include naming, but only the interpretation of given names.
    Two more were excluded because we wanted to limit the length of the experiment, and not squander precious experimental subjects.
    In the original study, each subject was also presented with six scenarios, which were randomly chosen from the set of 11 scenarios.
    In our version, all subjects saw the same six scenarios in the same order.

    In the original experiment, scenarios contained both questions that ask subjects to give names to variables and functions, and questions that ask them to interpret given names.
    Interpretation questions are not relevant for our research agenda.
    However, we decided not to remove them, so that our experiment will be as close as possible to the original experiment.

    \item Sending out invitations to participate in the experiment.
    The target audience was similar to that of the original study, and included professional developers and CS students.
    Potential participants were incentivized by a lottery for a gift card\anon{ of 300 NIS (about \$85)}{}.

    \item Presenting the scenarios to subjects on the Google Forms platform, and collection of the results.

    \item Downloading the results, and tabulating the distribution of the names given for each variable and function.
    This was then compared with the results obtained in the original study.
    Specifically we checked if there is a difference in the main results in terms of bias reduction, as reflected by the focus on the most popular answers.
\end{enumerate}

46 people responded to our questionnaire, of which 39 were students: 10 were studying for a BSc degree, 21 for an MSc, and 8 and for a PhD.
Nearly all were between 20--35 years of age, with nearly half between 25--30.
15 had no industrial programming experience, while 10 had 7 or more years of such experience.
This indicates, as is well-known, that many students work in parallel with their studies, and the division of experimental subjects into ``students'' and ``professionals'' is overly simplistic \cite{falessi18}.

\subsection{Results Normalization}

The last step of the execution of the experiment included a manual cleaning up of the results.
Specifically, the following changes where made to the raw results before they were analyzed:
\begin{itemize}
    \item Indications that the participant did not understand what to do were removed and treated as if no answer had been given.
    \item Indications of a namespace, such as \code{self.}, were removed, leaving only the bare name.
    \item In questions where a function signature was requested, the return type was removed, leaving only the function name.
    In the analysis reported here we also ignored any parameters that were given, as different participants used different numbers of parameters with different semantics.%, so it was impossible to consider them as alternative names for the same thing.
    \item In the elevator scenario we exchanged those cases where participants confused the variable referring to the current floor with the one referring to the destination floor, so as to avoid confusing mistakes in semantics with variations in naming (see below).
\end{itemize}

\section{Experimental results}
%=============================

\subsection{Understanding Emojis}
\label{sect:understand}

A premise of our work is that the experimental subjects understand the scenarios and questions presented to them using emojis.
Checking this completes the answer to \ref{rq:emojis}.

Two of the scenarios included questions that inadvertently check this directly.
These are questions about the interpretation of given names (a concept, a function name, and two parameters), where these names happen to be represented by emojis.
However, note that such questions were retained from the original experiment, and were not designed specifically to test the understanding of emojis.
The results of these questions are shown in Table \ref{tab:interpret}.
In one case 90\% were completely right and another 5\% were close.
In the other three cases about 3/4 were completely right, and a few more were close.
These results indicate that at least 80\% of the participants and in some cases perhaps up to 100\% understood the emojis.

\begin{table}\centering
    \caption{\sl Results of explicit interpretation questions.}
    \label{tab:interpret}
\begin{tabular}{|l|l|c|c|}
\hline
~Emojis~~ & Interpretation & $n$ & percent \\
\hline
\parbox[t][1.5ex]{0pt}{\raisebox{-2.0ex}{\pic{clock.png}{5mm}{-1.5mm}-\pic{money.png}{5mm}{-1.5mm}}}
& hourly wage & 33 & 77\% \\
& salary & 3 & 7\% \\
& other & 7 & 16\% \\
\hline
\parbox[t][1.5ex]{0pt}{\raisebox{-6.0ex}{\pic{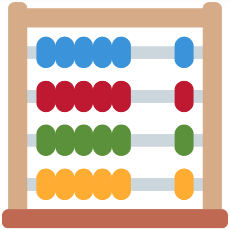}{10mm}{0mm}}}
& calculate & 37 & 90\% \\
& count & 2 & 5\% \\
& other & 2 & 5\% \\
\hline
\parbox[t][1.5ex]{0pt}{\raisebox{-6.0ex}{\pic{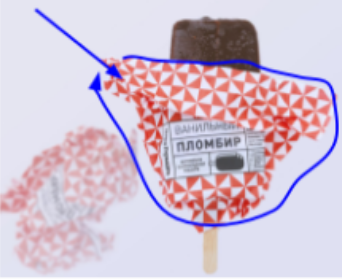}{12mm}{0mm}}}
& wrapper & 30 & 75\% \\
& paper & 3 & 8\% \\
& other & 5 & 18\% \\
\hline
\parbox[t][1.5ex]{0pt}{\raisebox{-5.0ex}{\pic{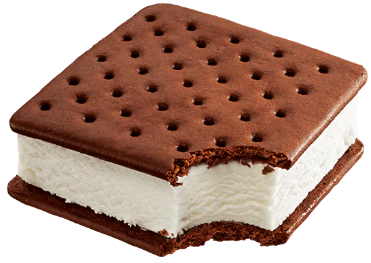}{10mm}{0mm}}}
& ice-cream sandwich & 30 & 73\% \\
& biscuit/cake & 7 & 17\% \\
& ice-cream & 4 & 10\% \\
\hline
\end{tabular}
\end{table}

But it is more important to see whether the subjects understood the scenarios and questions, not individual emojis.
Looking at the questions where subjects were required to give names to variable or functions,
in the vast majority of cases the names they gave indicated that they had understood the emojis correctly.
However, there was one case in which a sizeable fraction did not understand our intent.
The scenario was the well-known minesweeper game (which we referred to as a ``pastime'' in order to avoid the word ``game'').
One of the questions in this scenario was:
\begin{quote}
    Write a function signature for a function which receives \pic{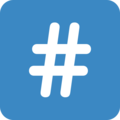}{4mm}{-1mm}\pic{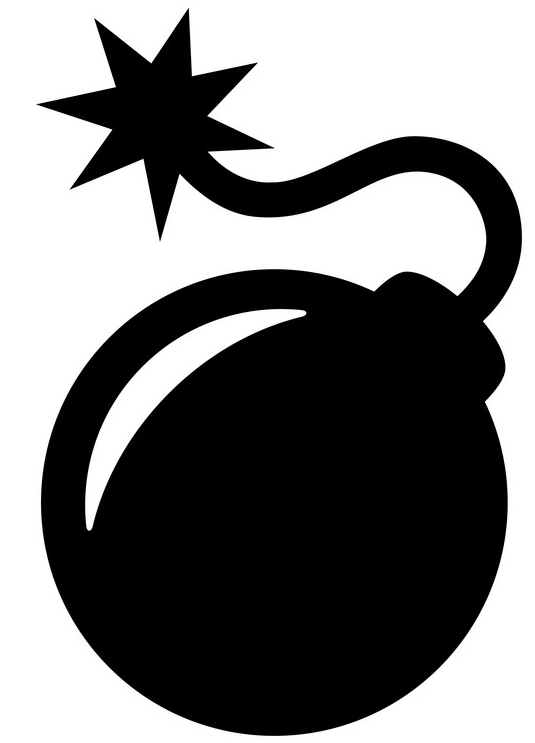}{4mm}{-1mm}\pic{bomb.png}{4mm}{-1mm}..., and \pic{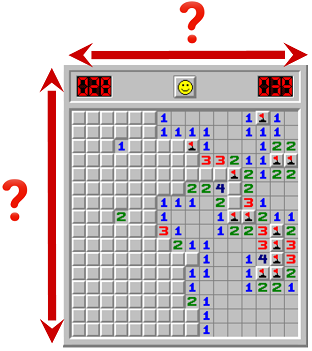}{15mm}{-3mm}\\
    and returns the \pic{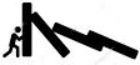}{12mm}{-1mm} of the pastime
\end{quote}
The icon \pic{push.png}{12mm}{-1mm} was supposed to reflect the difficulty of the game, and indeed 44\% of the subjects who answered this question understood this intent, giving names such as \code{get\_difficulty} and \code{calculate\_challenge\_level}.
But 15\% thought that we meant the construction or setup of a new game, giving names such as \code{createBoard}, 
and another 21\% thought we meant playing or solving the game, giving names such as \code{play\_mines} or \code{get\_user\_moves}.
In addition, 10\% explicitly indicated they did not understand what we meant.
% 39 answered. 17 difficulty, 6 build, 4 play, 4 solve, 4 don't know

Note, however, that not all mistakes or misunderstandings were due to emojis.
The most obvious example occurred in the elevator scenario.
One of the questions in this scenario included the following code:
\code{\begin{tabbing}
    xxxx\=xxxx\= \+ \kill
    if (var1 > var2)\\
    \> direction = "Up"\\
    \> var3 = var1 - var2\\
    \> \pic{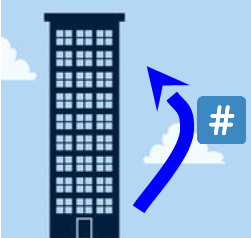}{15mm}{-4mm}(var3)\\[2mm]
    if (var1 < var2)\\
    \> direction = "Down"\\
    \> var3 = var2 - var1\\
    \> \pic{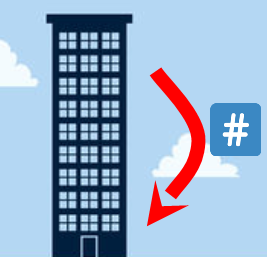}{15mm}{-4mm}(var3)
\end{tabbing}}
(where the icons replace the function names \code{goUp} and \code{goDown} from the original version, and the subjects knew that they represent function names but of course not the names themselves).
The question then was to give better names to \code{var1}, \code{var2}, and \code{var3}.
The answers revealed that 66\% of those who answered this question understood that \code{var1} represents the destination floor and \code{var2} represents the current floor, but 26\% got confused and mistakenly thought it was the other way around.
Consequently, just counting mistakes is not a good way to assess the understanding of emojis.

% moved up
\begin{table*}\centering
\caption{\label{tab:2hit}\sl
Results of name reuse in all versions of questions concerned with giving names.
Note that these results are for the complete names, not for concepts; using an abbreviation or a different word order is counted as a different name.
\emph{$N$}: number of answers to this question;
\emph{dif}: number of different names given;
\emph{div=dif/$N$}: diversity of names;
\emph{max}: maximal answers giving the same name;
\emph{Pmax=max/$N$}: probability of most popular name (focus);
\emph{P2hit}: probability of two participants using the same name.
}
\begin{tabular}{llcccccc}
\hline
\emph{Scenario} & \emph{Question} & \emph{$N$} & \emph{dif} & \emph{div} & \emph{max} & \emph{Pmax} & \emph{P2hit} \\
\hline
% Q3_2.txt
Candy   & constant specifying work hours per week
        & 44    & 40    & 0.909 & 2     & 0.045 & 0.0268 \\

% Q3_3.txt
factory & variable holding hourly wage during overtime
        & 42    & 37    & 0.880 & 4     & 0.095 & 0.0328 \\
\hline

% Q6_2.txt
Elevator        & variable with requested floor
        & 40    & 27    & 0.675 & 9     & 0.225 & 0.0775 \\

% Q6_3.txt
        & variable with current elevator location
        & 40    & 17    & 0.425 & 17    & 0.425 & 0.2124 \\

% Q6_4.txt
        & variable with number of floors to move
        & 39    & 26    & 0.666 & 5     & 0.128 & 0.0571 \\

% Q6_5.txt
        & variable with state of elevator doors
        & 40    & 16    & 0.400 & 15    & 0.375 & 0.1787 \\
\hline

% Q1_1.txt
File    & field in file object describing file size
        & 46    & 9     & 0.195 & 30    & 0.652 & 0.4669 \\

% Q1_2.txt
system  & function checking if there is enough disk space to extend a file
        & 46    & 39    & 0.847 & 3     & 0.065 & 0.0302 \\
\hline

% Q2_2.txt
Mine-   & function calculating game's difficulty level
        & 35    & 27    & 0.771 & 4     & 0.114 & 0.0497 \\

% Q2_3.txt
sweeper & variable with game's time
        & 43    & 25    & 0.581 & 6     & 0.139 & 0.0665 \\

% Q2_4.txt
        & data structure indicating mine or number of adjacent mines
        & 39    & 32    & 0.820 & 6     & 0.153 & 0.0479 \\
\hline

% Q5_1.txt
Benefits & constant with value 4 (max benefits per month)
        & 41    & 26    & 0.634 & 8     & 0.195 & 0.0719 \\

% Q5_2.txt
card    & constant with value 2000 (shekels per benefits point)
        & 41    & 37    & 0.902 & 5     & 0.121 & 0.0362 \\

% Q5_3.txt
        & variable with entitled benefits this month
        & 41    & 41    & 1.000 & 1     & 0.024 & 0.0243 \\

% Q5_4.txt
        & function checking if balance of benefits is positive
        & 41    & 41    & 1.000 & 1     & 0.024 & 0.0243 \\
\hline

% Q4_1.txt
Ice cream & function calculating how many sandwiches can be produced
        & 41    & 40    & 0.975 & 2     & 0.048 & 0.0255 \\

% median diversity is 0.820
% median Pmax is 0.128
% median P2hit is 0.0497
\hline
\end{tabular}
\end{table*}

\subsection{Name Choice}
\label{sect:name-choice}

The goal of our work was to investigate whether using emojis in lieu of key words in scenarios and questions can reduce the accessibility bias caused by using such words.
One aspect of this issue is whether participants tended to use the same names --- presumably due to being influenced by the scenario description and the question, or perhaps they tended to come up with totally different names.

Feitelson et al.\ defined two related metrics for the distribution of names \cite{feitelson:names}:
\begin{itemize}
\item The degree of \textbf{focus} on using a particular name.
This was quantified as the fraction of times that the most popular name was used.
\item The \textbf{diversity} of names used.
This was quantified as the quotient of the number of different names given divided by the total number of responses received.
\end{itemize}

The results for the 16 questions concerned with naming variables and functions in the 6 scenarios we used are tabulated in Table \ref{tab:2hit}.
This includes both the raw results and the quotients as defined above.
In addition, the last column presents an estimate of the probability that two different developers would use the same name.
Following Feitelson et al.\ this is calculated as \cite{feitelson:names}
\begin{equation}\label{eq:p2hit}
    P2hit = \sum_{i=1}^k p_i^2
\end{equation}
where we observe $k$ distinct names, and $p_i$ denotes the probability of choosing name $i$ (as estimated by its relative popularity).

\begin{figure}\centering
\includegraphics[width=0.85\columnwidth]{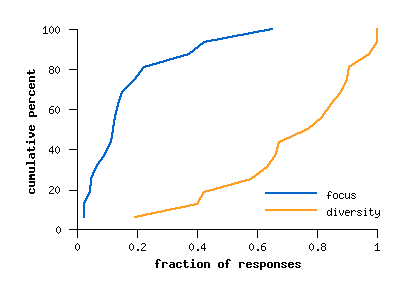}
\vspace*{-4mm}
\caption{\label{fig:focus-divers}\sl
Cumulative distribution of focus and diversity of names given in the experiment questions.}
\end{figure}

Fig.\ \ref{fig:focus-divers} shows the distribution of the focus and diversity across the 16 questions.
The median focus was 0.13, and in 75\% of the cases the focus was less than 0.2.
The maximal focus was 0.65, which occurred when \code{size} was used to name a field containing the size of a file in a data structure representing the file.
This was also the instance with the highest focus in the original study.

The median diversity was 0.8, and in 75\% of the cases the diversity was higher than 0.6.
In two cases the diversity was 1: all the names were different from each other.
The lowest diversity, 0.2, occurred in the same case as the highest focus, and indeed there is a strong negative correlation between the focus and the diversity with $\rho=-$0.93.

The median estimated probability for choosing the same name was 0.049, lower than the 0.069 found in the original study.
In 75\% of the cases the estimate was lower than 0.072.
The highest estimate, 0.47, occurred again in the case of the \code{size} field.
The next highest estimate was less than half this value.

\subsection{Accessibility Bias}

Our main goal is to see whether using emojis can reduce the accessibility bias, as reflected in \ref{rq:accessibility}.
We therefore conducted an in-depth investigation of the concepts and words used by the experiment participants when answering each question.
Feitelson et al.\ \cite{feitelson:names} present a tool used to manually identify the concepts embedded in each name and the words representing each concept.
We used this tool to do the same for our data, and compare with the results of Feitelson et al.\ \cite{feitelson:names} (which were not presented in detail in their paper).

\begin{table*}\centering
    \caption{\sl Concept usage in the different questions of the experiment.
    Resuls based on emojis (Emj) are ours.
    Results about English and Hebrew versions are from the experimental materials of Feitelson et al.\ \cite{feitelson:names}.}
    \label{tab:concepts}
    \begin{tabular}{llccccl}

\hline
 & \bf Concept & \multicolumn{2}{l}{\bf Importance} & \bf Focus & \bf Div. & \bf Dominant words \\

% Q54
\hline
\multicolumn{7}{l}{\it Candy: constant specifying work hours per week} \\
\hline
Eng & hours	& 0.78 & M &  0.90 & 0.09 & hours (90\%)\\
 & full	& 0.36 & I &  1 & 0.06 & full\_time (100\%)\\
 & threshold	& 0.34 & I &  0.21 & 0.50 & capacity (21\%), min (21\%), threshold (21\%)\\
 & week	& 0.34 & I &  0.85 & 0.21 & weekly (85\%)\\
 & job	& 0.21 & - &  0.44 & 0.44 & work (44\%)\\
% & rate	& 0.19 & - &  0.25 & 0.87 & \\
 & normal	& 0.19 & - &  0.37 & 0.62 & base (37\%)\\
% & number	& 0.02 & - &  1 & 1 & \\
\hline
% Q44
Heb & hours	& 0.72 & M &  0.96 & 0.06 & hours (96\%)\\
 & full	& 0.51 & I &  0.50 & 0.13 & full\_time (50\%), full (45\%)\\
 & job	& 0.46 & I &  0.40 & 0.30 & job (40\%), work (30\%), overtime (15\%)\\
 & week	& 0.32 & I &  0.42 & 0.35 & per\_week (42\%), week (21\%), weekly (21\%)\\
 & threshold	& 0.20 & - &  0.44 & 0.44 & max (44\%), threshold (33\%)\\
 & normal	& 0.16 & - &  0.42 & 0.42 & base (42\%)\\
% & rate	& 0.13 & - &  0.83 & 0.33 & rate (83\%)\\
\hline
% Q3_2
Emj & hours	& 0.78 & M &  0.62 & 0.06 & hours (62\%), time (37\%)\\
 & weekly	& 0.54 & I &  0.40 & 0.15 & week (40\%), weekly (40\%), per\_week (20\%)\\
 & regular	& 0.48 & I &  0.22 & 0.72 & base (22\%)\\
 & work	& 0.45 & I &  0.88 & 0.17 & work (88\%)\\
% & length	& 0.29 & - &  0.18 & 0.72 & \\
\hline

% Q55
\hline
\multicolumn{7}{l}{\it Candy: variable holding hourly wage during overtime} \\
\hline
Eng & pay\_rate	& 1 & D &  0.39 & 0.14 & wage (39\%), rate (34\%), pay (12\%), salary (7\%)\\
 & overtime	& 0.82 & M &  0.85 & 0.14 & overtime (85\%)\\
% & hours	& 0.26 & - &  0.81 & 0.27 & hourly (81\%)\\
 & add	& 0.12 & - &  0.40 & 0.80 & \\
\hline
% Q45
Heb & overtime	& 0.83 & M &  0.27 & 0.25 & extra\_hours (27\%), overtime (27\%), extra (11\%), extra\_time (11\%)\\%, additional\_hours (8\%)\\
 & pay\_rate	& 0.74 & M &  0.65 & 0.25 & rate (65\%), payment (9\%)\\
 & add	& 0.39 & I &  0.41 & 0.29 & addition (41\%), bonus (41\%)\\
% & hour	& 0.25 & - &  0.45 & 0.36 & per\_hour (45\%), hourly (27\%)\\
\hline
% Q3_3
Emj & salary	& 0.85 & M &  0.35 & 0.26 & salary (35\%), rate (20\%), wage (11\%), money (8\%), payment (8\%)\\
 & overtime	& 0.70 & M &  0.60 & 0.28 & overtime (60\%), extra\_hours (17\%)\\
% & hourly	& 0.27 & - &  0.54 & 0.27 & hourly (54\%), time (27\%)\\
 & extra	& 0.25 & - &  0.30 & 0.60 & bonus (30\%), increase (30\%)\\
\hline

% Q123
\hline
\multicolumn{7}{l}{\it Elevator: variable with requested floor} \\
\hline
Eng & destination	& 1 & D &  0.33 & 0.43 & destination (33\%), requested (10\%), target (10\%)\\
 & floor	& 0.87 & M &  0.91 & 0.08 & floor (91\%)\\
\hline
% Q130
Heb & destination	& 1 & D &  0.35 & 0.30 & destination (35\%), target (22\%), requested (10\%), desired (7\%)\\
 & floor	& 0.90 & D &  0.88 & 0.11 & floor (88\%)\\
\hline
% Q6_2
Emj & destination	& 0.97 & D &  0.30 & 0.35 & target (30\%), destination (23\%), requested (7\%)\\
 & floor	& 0.80 & M &  0.90 & 0.09 & floor (90\%)\\
% & number	& 0.02 & - &  1 & 1 & \\
\hline

% Q124
\hline
\multicolumn{7}{l}{\it Elevator: variable with current elevator location} \\
\hline
Eng & current	& 1 & D &  0.91 & 0.11 & current (91\%)\\
 & floor	& 0.97 & D &  0.90 & 0.12 & floor (90\%)\\
\hline
% Q131
Heb & current	& 0.95 & D &  0.87 & 0.12 & current (87\%)\\
 & floor	& 0.92 & D &  0.84 & 0.10 & floor (84\%), level (7\%)\\
% & elevator	& 0.02 & - &  1 & 1 & \\
\hline
% Q6_3
Emj & current	& 0.97 & D &  0.81 & 0.15 & current (81\%)\\
 & floor	& 0.87 & M &  0.88 & 0.11 & floor (88\%)\\
% & number	& 0.02 & - &  1 & 1 & \\
\hline

% Q125
\hline
\multicolumn{7}{l}{\it Elevator: variable with number of floors to move} \\
\hline
Eng & floor	& 0.87 & M &  0.91 & 0.11 & floor (91\%)\\
 & difference	& 0.58 & I &  0.47 & 0.26 & difference (47\%), delta (21\%), distance (17\%)\\
 & move	& 0.28 & - &  0.72 & 0.27 & move (72\%)\\
% & number	& 0.28 & - &  0.9 & 0.18 & number (90\%)\\
\hline
% Q132
Heb & floor	& 0.76 & M &  0.91 & 0.08 & floor (91\%)\\
 & difference	& 0.46 & I &  0.36 & 0.31 & difference (36\%), delta (27\%), gap (13\%)\\
 & move	& 0.38 & I &  0.61 & 0.16 & move (61\%), to\_go (22\%), travel (16\%)\\
% & number	& 0.27 & - &  0.76 & 0.3 & number (76\%)\\
% & destination	& 0.02 & - &  1 & 1 & \\
\hline
% Q6_4
Emj & floors	& 0.79 & M &  0.96 & 0.06 & floors (96\%)\\
 & move	& 0.48 & I &  0.57 & 0.21 & move (57\%), travel (26\%)\\
 & difference	& 0.48 & I &  0.52 & 0.31 & difference (52\%), distance (15\%), delta (15\%)\\
% & number	& 0.2 & - &  0.87 & 0.25 & number (87\%)\\
\hline

% Q127
\hline
\multicolumn{7}{l}{\it Elevator: variable with state of elevator doors} \\
\hline
Eng & door	& 0.67 & I &  1 & 0.03 & door (100\%)\\
 & state	& 0.50 & I &  1 & 0.05 & state (100\%)\\
 & open	& 0.47 & I &  0.89 & 0.10 & open (89\%)\\
 & current	& 0.47 & I &  0.84 & 0.10 & is (84\%), current (15\%)\\
% & elevator	& 0.1 & - &  1 & 0.25 & elevator (100\%)\\
\hline
% Q134
Heb & door	& 0.79 & M &  1 & 0.02 & door (100\%)\\
 & current	& 0.55 & I &  0.96 & 0.07 & is (96\%)\\
 & open	& 0.53 & I &  0.84 & 0.07 & open (84\%), close (15\%)\\
 & state	& 0.40 & I &  0.65 & 0.10 & state (65\%), status (35\%)\\
% & elevator	& 0.02 & - &  1 & 1 & \\
\hline
% Q6_5
Emj & open	& 0.72 & M &  0.86 & 0.06 & open (86\%), closed (13\%)\\
 & door	& 0.60 & I &  1 & 0.04 & door (100\%)\\
 & current	& 0.55 & I &  1 & 0.04 & is (100\%)\\
 & state	& 0.27 & - &  0.63 & 0.27 & state (63\%), status (27\%)\\
% & elevator	& 0.02 & - &  1 & 1 & \\
\hline

    \end{tabular}
\end{table*}

\begin{table*}\centering
    \addtocounter{table}{-1}
    \caption{\sl Concept usage in the different questions of the experiment (continued).}
    \begin{tabular}{llccccl}

\hline
 & \bf Concept & \multicolumn{2}{l}{\bf Importance} & \bf Focus & \bf Div. & \bf Dominant words \\

% Q99
\hline
\multicolumn{7}{l}{\it Files: field in file object describing file size} \\
\hline
Eng & size	& 1 & D &  0.97 & 0.04 & size (97\%)\\
 & file	& 0.45 & I &  1 & 0.05 & file (100\%)\\
% & byte	& 0.04 & - &  1 & 0.5 & \\
% & field	& 0.02 & - &  1 & 1 & \\
\hline
% Q105
Heb & size	& 1 & D &  1 & 0.01 & size (100\%)\\
 & file	& 0.52 & I &  1 & 0.03 & file (100\%)\\
% & byte	& 0.05 & - &  0.66 & 0.66 & \\
\hline
% Q1_1
Emj & size	& 0.93 & D &  1 & 0.02 & size (100\%)\\
 & file	& 0.28 & - &  0.92 & 0.15 & file (92\%)\\
% & storage	& 0.08 & - &  0.5 & 0.75 & \\
\hline

% Q37
\hline
\multicolumn{7}{l}{\it Minesweeper: variable with game's time} \\
\hline
Eng & time	& 0.93 & D &  0.81 & 0.06 & time (81\%), timer (13\%)\\
 & game	& 0.51 & I &  0.91 & 0.12 & game (91\%)\\
 & length	& 0.21 & - &  0.40 & 0.50 & elapsed (40\%), duration (30\%)\\
% & apply	& 0.08 & - &  0.5 & 0.5 & \\
% & counter	& 0.06 & - &  1 & 0.33 & counter (100\%)\\
\hline
% Q31
Heb & time	& 0.86 & M &  0.67 & 0.13 & time (67\%), timer (18\%)\\
 & length	& 0.40 & I &  0.3 & 0.45 & elapsed (30\%), duration (25\%)\\
 & game	& 0.38 & I &  0.89 & 0.10 & game (89\%)\\
% & apply	& 0.12 & - &  0.5 & 0.66 & current (50\%)\\
\hline
% Q2_3
Emj & time	& 0.97 & D &  0.61 & 0.14 & time (61\%), timer (11\%), seconds (11\%), clock (9\%)\\
 & length	& 0.34 & I &  0.40 & 0.40 & elapsed (40\%), passed (33\%)\\
% & apply	& 0.13 & - &  0.66 & 0.5 & remain (66\%)\\
 & game	& 0.09 & - &  0.50 & 0.75 & \\
\hline

% Q36
\hline
\multicolumn{7}{l}{\it Minesweeper: data structure indicating mine or number of adjacent mines} \\
\hline
Eng & board	& 0.59 & I &  0.34 & 0.26 & board (34\%), grid (30\%), map (15\%)\\
 & mines	& 0.40 & I &  1 & 0.05 & mines (100\%)\\
 & cell	& 0.22 & - &  0.50 & 0.40 & cell (50\%), square (30\%)\\
 & state	& 0.20 & - &  0.33 & 0.55 & data (33\%), value (33\%)\\
% & game	& 0.15 & - &  0.85 & 0.28 & game (85\%)\\
% & adjacent	& 0.06 & - &  1 & 0.33 & adjacent (100\%)\\
 & number	& 0.04 & - &  1 & 0.50 & \\
% & action	& 0.04 & - &  0.5 & 1 & \\
\hline
% Q32
Heb & board	& 0.53 & I &  0.50 & 0.28 & board (50\%), field (10\%), map (10\%), matrix (10\%)\\
 & mines	& 0.46 & I &  0.83 & 0.12 & mines (83\%), bomb (12\%)\\
 & state	& 0.30 & I &  0.18 & 0.62 & values (18\%), info (18\%)\\
 & cell	& 0.28 & - &  0.60 & 0.26 & cell (60\%), square (26\%)\\
% & game	& 0.11 & - &  0.83 & 0.33 & game (83\%)\\
 & number	& 0.07 & - &  0.75 & 0.50 & number (75\%)\\
% & adjacent	& 0.03 & - &  0.5 & 1 & \\
\hline
% Q2_4
Emj & count	& 0.44 & I &  0.52 & 0.35 & values (52\%), number (17\%)\\
 & board	& 0.39 & I &  0.66 & 0.33 & board (66\%)\\
 & tiles	& 0.36 & I &  0.50 & 0.35 & cell (50\%), tile (28\%)\\
 & mines	& 0.23 & - &  0.55 & 0.33 & bomb (55\%), mine (33\%)\\
% & near	& 0.21 & - &  0.25 & 0.62 & \\
% & game	& 0.02 & - &  1 & 1 & \\
\hline

% Q136
\hline
\multicolumn{7}{l}{\it Card benefits: constant with value 4 (max benefits per month)} \\
\hline
Eng & benefit	& 0.94 & D &  0.98 & 0.03 & benefit (98\%)\\
 & max	& 0.88 & M &  0.89 & 0.06 & max (89\%), limit (8\%)\\
 & month	& 0.38 & I &  0.71 & 0.14 & per\_month (71\%), monthly (19\%)\\
% & number	& 0.27 & - &  0.66 & 0.2 & number (66\%), accumulation (26\%)\\
% & allowed	& 0.03 & - &  0.5 & 1 & \\
% & customer	& 0.01 & - &  1 & 1 & \\
\hline
% Q141
Heb & benefit	& 0.97 & D &  0.28 & 0.28 & pinuk (28\%), treat (21\%), gift (14\%), benefit (11\%)\\
 & max	& 0.88 & M &  0.92 & 0.07 & max (92\%)\\
 & month	& 0.41 & I &  0.55 & 0.16 & per\_month (55\%), monthly (27\%), month (16\%)\\
% & number	& 0.11 & - &  0.8 & 0.4 & number (80\%)\\
\hline
% Q5_1
Emj & points	& 0.92 & D &  0.62 & 0.32 & diamonds (62\%), points (10\%)\\
 & max	& 0.92 & D &  0.91 & 0.08 & max (91\%)\\
 & per\_month	& 0.37 & I &  0.66 & 0.20 & per\_month (66\%), monthly (20\%)\\
% & count	& 0.2 & - &  0.62 & 0.25 & accumulated (62\%), number (37\%)\\
\hline

% Q137
\hline
\multicolumn{7}{l}{\it Card benefits: constant with value 2000 (shekels per benefits point)} \\
\hline
Eng & amount	& 0.78 & M &  0.27 & 0.39 & ils (27\%), value (11\%), amount (9\%), sum (6\%), money (6\%)\\
 & benefit	& 0.78 & M &  0.97 & 0.04 & benefit (97\%)\\
 & threshold	& 0.32 & I &  0.27 & 0.55 & min (27\%), threshold (22\%)\\
% & ratio	& 0.12 & - &  0.28 & 0.71 & \\
% & month	& 0.01 & - &  1 & 1 & \\
\hline
% Q142
Heb & benefit	& 0.90 & D &  0.30 & 0.30 & pinuk (30\%), treat (23\%), gift (12\%), benefit (7\%)\\
 & amount	& 0.79 & M &  0.20 & 0.41 & money (20\%), price (14\%), cost (11\%), amount (8\%), spending (8\%)\\
 & threshold	& 0.34 & I &  0.33 & 0.53 & threshold (33\%), min (26\%)\\
% & increment	& 0.06 & - &  0.33 & 1 & \\
% & ratio	& 0.04 & - &  1 & 0.5 & \\
% & month	& 0.02 & - &  1 & 1 & \\
\hline
% Q5_2
Emj & points	& 0.87 & M &  0.64 & 0.29 & diamond (64\%), point (8\%)\\
 & money	& 0.84 & M &  0.24 & 0.36 & value (24\%), dollar (15\%), money (15\%), shekel (12\%), spending (9\%)\\
 & ratio	& 0.74 & M &  0.37 & 0.34 & per (37\%), rate (13\%), for (13\%), to (10\%)\\
\hline

% Q138
\hline
\multicolumn{7}{l}{\it Card benefits: variable with entitled benefits this month} \\
\hline
Eng & benefit	& 0.94 & D &  0.98 & 0.03 & benefit (98\%)\\
 & number	& 0.49 & I &  0.70 & 0.14 & number (70\%), count (22\%)\\
 & current	& 0.41 & I &  1 & 0.04 & current (100\%)\\
 & month	& 0.3 & I &  0.94 & 0.11 & month (94\%)\\
 & available	& 0.3 & I &  0.58 & 0.41 & entitlement (58\%)\\
% & customer	& 0.07 & - &  0.75 & 0.5 & client (75\%)\\
\hline
% Q143
Heb & benefit	& 0.97 & D &  0.28 & 0.26 & pinuk (28\%), treat (21\%), gift (16\%), benefit (9\%), bonus (7\%)\\
 & number	& 0.37 & I &  0.81 & 0.25 & number (81\%)\\
 & current	& 0.34 & I &  1 & 0.06 & current (100\%)\\
 & available	& 0.30 & I &  0.38 & 0.69 & available (38\%)\\
 & month	& 0.25 & - &  0.72 & 0.27 & month (72\%)\\
% & customer	& 0.11 & - &  0.4 & 0.8 & \\
% & threshold	& 0.04 & - &  0.5 & 1 & \\
\hline
% Q5_3
Emj & points	& 0.97 & D &  0.65 & 0.27 & diamonds (65\%), points (12\%)\\
 & have	& 0.53 & I &  0.18 & 0.50 & earned (18\%), entitled (18\%), available (13\%), accumulated (13\%)\\
 & time	& 0.39 & I &  0.37 & 0.31 & current (37\%), current\_month (25\%), monthly (18\%)\\
 & number	& 0.36 & I &  0.80 & 0.26 & number (80\%)\\
\hline

    \end{tabular}
\end{table*}

Table \ref{tab:concepts} shows the identified concepts used by participants in our experiment (based on emoji descriptions) compared with the concepts used by participants in the original experiment (based on descriptions in English and Hebrew).
We define concept importance levels as follows:
\begin{itemize}
    \item \textbf{D}ominant = appears in at least 90\% of the names
    \item \textbf{M}ajor = appears in at least 70\% of the names
    \item \textbf{I}mportant = appears in at least 30\% of the names
    \item unimportant = appears in less than 30\% of the names
\end{itemize}
Unimportant concepts are shown only if they are important in another experimental setting.
For each concept, we also calculate focus and diversity of its words, and show the top words used to represent this concept.

\begin{figure}\centering
\includegraphics[width=0.85\columnwidth]{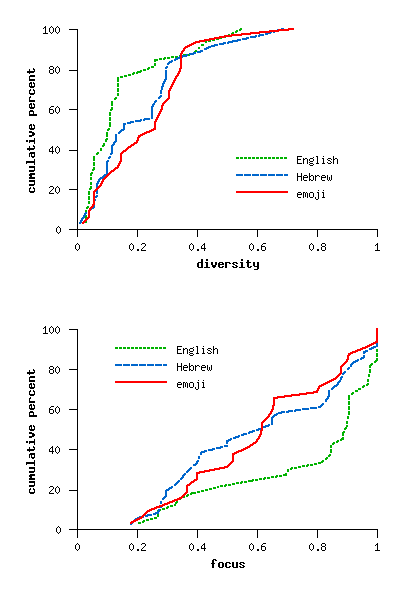}
\vspace*{-4mm}
\caption{\label{fig:focus-divers-cmp}\sl
Comparison of the cumulative distributions of focus and diversity of important concepts, for English and Hebrew descriptions from Feitelson et al.\ \cite{feitelson:names} and emoji descriptions from our experiment.}
\end{figure}

Looking at the results, three classes of questions can be identified.
One class contains mainly questions that have an obvious dominant answer.
For example, when asking about the field containing the file size in a file system context, practically all participants used the word ``size'' and about half also used ``file''.
In the question about overtime wages in the candy factory scenario, a mix of different words were used, but the top concepts were ``overtime'' and ``salary''.
In these cases the presentation style (English, Hebrew, or emojis) does not make a big difference.

Another class of questions are those where using emojis \emph{did} make a difference.
In some cases different concepts were used:
\begin{itemize}
    \item In the question concerning the variable for holding the playing time of the minesweeper game, the top concepts in the original experiment were ``time'', ``game'', and ``length''.
    In our emoji version, ``game'' was very seldom used leaving only two important concepts.
    \item For the data structure that maps cells to mines, in the original experiment ``board'' and ``mine'' were the top concepts, the concept of ``cell'' was used much less, and the concept of ``number'' was hardly used.
    In the emoji version ``number'' was the top concept, followed by ``board'' and ``tiles'' (cells).
\end{itemize}
In other cases, different words were used.
For example, Around half the participants included the concept of ``full-time'' in the constant specifying the working hours per week in the candy factory scenario.
In the original experiment this was an extremely focused concept.
With emojis it was extremely diverse.

The third and final class is where the emoji used caused an unintended accessibility effect of its own.
We had one such example: the credit card scenario, where benefit points accrued by charging expenses to the credit card were represented by \pic{diamond.png}{6mm}{-2mm}.
This choice caused more than half of the participants to use the word ``diamond''.
The effect was not as strong as the English version of the original experiment (where ``benefits'' was used by nearly everyone), but stronger than in the Hebrew version (where the transliteration of the Hebrew word ``pinuk'' was used by about 30\%).

Fig.\ \ref{fig:focus-divers-cmp} shows the cumulative distribution of the focus and diversity of word use in all the important concepts (importance$\ge$0.3) from all the questions.
The results when using emojis is compared with the results from the original experiment of Feitelson et al.\ \cite{feitelson:names}.
It is easy to see that when using English the focus is higher and the diversity is lower.
Emojis are similar to using Hebrew, perhaps even leading to slightly higher diversity.

\section{Threats to validity}
%============================

\paragraph{Construct Validity}
Our measurements may not measure the right thing for several reasons, e.g.\ if the questions we used in the experiment were too simple or the text was not rich enough.
In the industry specifications are much longer and exhausting to read.
Short descriptions like ours may therefore guide developers to the same variable name, even after the text was translated to emojis.

It may also be that different cultures or genders will have different perceptions of specific emojis, leading to a bias.
For instance, does \pic{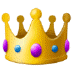}{5mm}{-1mm} represent the word ``king'', ``queen'', or just ``crown''?
Moreover, there is a danger that emojis will be misunderstood, as the vocabulary of emojis is rather limited.\\

\paragraph{Internal Validity}
Our conclusions may not follow from the experiments for several reasons.
Our questionnaire response rate was somewhat lower than the response rate of the original study, which might have an effect on the results.

The translation to emojis may be problematic.
The vocabulary of emojis is limited and does not include may concepts and nuances.
In some cases we used somewhat distorted language, or even modified the description and questions, so as to avoid explicit key words that might cause an accessibility bias.
This may affect understanding, and especially limit the comparability with the results of the original experiment.\\

\paragraph{External Validity}
Our research might not be generalizable.
Our experimental subjects were mostly from academia\anon{ and specifically from the Hebrew University}{}, which might cause a bias:
they mostly underwent university education, which might be different from college education or self-education;
many had limited or no industrial experience;
they have the free time for participating in an experiment;
and they might participate because they know us personally, further reducing their representativeness.

\section{Discussion and Conclusions}
%===================================

We have replicated a study on variable and function naming by Feitelson et al.\ \cite{feitelson:names}.
In the original study the accessibility bias was reduced by employing bilingual experimental subjects, and providing scenario descriptions in Hebrew rather than in English.
In our replications the descriptions were in English, but key words were replaced by emojis.
The results indicate that this reduced the accessibility bias to a similar degree as using a foreign language.
As a result names tended to have a more diverse use of words.

However, in some cases not only the words but also the concepts embedded in names when the descriptions used emojis were somewhat different from the concepts when the descriptions were verbal.
Additional work is therefore needed to better understand how developers select the concepts to embed in names.

An important observation, both in the original experiment of Feitelson et al.\ \cite{feitelson:names} and in our replication using emojis, is the focus-diversity dichotomy.
This occurs both at the level of complete names and at the level of words used to represent an individual concept within names.
Focus means that a single name or word dominates, and few others are used.
Diversity means that many different names or words are used, and none of them dominates.
The results indicate that in some cases we find high focus and low diversity, while in others the focus is low and the diversity high.

In terms of program comprehension research, focused situations are probably not very interesting.
The focus testifies that all participants understood the situation in the same way and expressed themselves in the same way.
It stands to reason that the chosen names will be easily understood.

The diverse situations, on the other hand, are those that deserve further study.
The diversity in names and words testifies that there may be different ways to understand the situation.
And even if not, using different words implies different semantic nuances.
Thus choosing different names and words may cause misunderstandings and confusion.

Emojis can help studying diverse naming by helping to eliminate, or at least decrease, the accessibility bias.
By using emojis we can avoid the use of specific words, and thereby avoid implanting specific ideas in the minds of experimental subjects.
However, emojis are not a panacea.
Problems that can occur with emojis can lead either to high focus or to high diversity, in both cases threatening the experimental validity.
\begin{itemize}
\item High focus can be a sign of a crisp, well-defined situation as noted above.
However, it can also reflect a problematic description using emojis:
\begin{itemize}
    \item Using too obvious emojis with a specific obvious meaning.
    This is especially harmful if this meaning is not exactly the intended one.
    \item Important key words were not translated, and the description is still very verbal.
    In other words, an explicit accessibility bias exists.
\end{itemize}
In these situations, the experiment does not expose the potential diversity of names and the misunderstandings that they may cause.
\item High diversity can also be a result of problems with the emoji representation:
\begin{itemize}
    \item When the emoji is ambiguous, and can reasonably be interpreted in different ways.
    \item When the whole descriptions becomes vague and puzzling, due to modifications made as part of the emoji translation.
\end{itemize}
In these situations the desired functionality of the variable becomes less understandable, rendering experiments concerning its naming invalid.
\end{itemize}
Both these cases show the importance a running a pilot on the emojis descriptions.
By conducting a pilot with an adequate debriefing of the participants one can reduce the danger of both over-specific emojis and ambiguous ones.

To conclude, we believe that non-verbal expressions such as emojis should be considered as a tool that improves research in naming.
Descriptions with emojis are similarly expressive as verbal text, and emojis are generally well understood.
In addition, emojis can reduce the accessibility bias.
And this approach has the advantage of not requiring experimental subjects who are fluent in a foreign language, making it suitable for English speaking experimental subjects.

\section*{Experimental Materials}

The experimental materials, including survey questions, responses, analysis scripts, and results, are available using DOI \url{https://doi.org/10.5281/zenodo.4603985}.

%\IEEEtriggeratref{19}
\bibliographystyle{myabbrv}
\bibliography{abbrv,se}

\end{document}